\documentclass{article}
\usepackage{graphicx,amssymb,amsmath,times,cite,psfrag,a4}
\begin{document}
\title{Primary proton spectrum in the 
energy range $\mathbf{5}$--$\mathbf{10^3}$~TeV from the sea level muon 
spectrum} 
\author {A.\,A. Lagutin, A.\,G. Tyumentsev, A.\,V. Yushkov} 
\date{\small Altai State University, Lenin Ave. 61, Barnaul 656049, Russia\\
e-mail: yushkov@theory.dcn-asu.ru}
\maketitle

\begin{abstract}
Primary proton spectrum in the energy range $5-10^3$ TeV is reconstructed from 
the sea level muon spectrum with the use of QGSJET01 and SYBILL2.1 interaction 
models. Heavier nuclei are taken in accordance with the direct measurements 
data, 100\% uncertainty in helium flux is accounted for. The obtained proton 
intensity strongly contradicts to the available data of balloon experiments, 
exceeding them at the least by 100\% for QGSJET01. This discrepancy is due to 
the combined effect of primary nucleon flux underestimation in the direct 
measurements and incorrect description of extensive air shower development. In 
the latter case it is required earlier shower development and harder spectra of 
secondary pions and kaons in comparison with QGSJET01. This conclusion is in 
agreement with the obtained by the KASCADE group on the basis of events rate 
study. 
\end{abstract}

\section*{Primary proton spectrum from the different EAS observables}

Recently it was shown~\cite{ya2004,unger_ecrs2004}, that the use of direct data 
on primary cosmic rays (PCR) spectra and hadronic interaction models, included 
in CORSIKA, leads to significantly underrated, in comparison with the 
measurements, sea level muon flux for $E_\mu>100$ GeV. The discrepancy takes 
place already for energies well below the ``knee'' ($E_\text{PCR}\lesssim100$ 
TeV), where behavior of primary nucleon flux and hadronic interaction 
cross-sections seems to be rather reliably established. Attempts to explain the 
lack of high-energy muons by errors in EAS 
simulation~\cite{unger_ecrs2004,engel_pylos} touch only one side of the 
problem, since direct data on PCR spectra are far from being considered as 
reference values. The emulsion chamber (EC) technique, applied in balloon 
experiments, is extremely labor consuming and 
sophisticated~\cite{grigorov1973_eng,jacee1986,runjob_eng}, and final results 
(PCR fluxes) are sensitive to many factors: from purely instrumental to the 
choice of hadronic generator. As a consequence, these experiments have limited 
energy resolution and disagree on the fluxes of nuclei with $Z\geq2$. 

The fact, that SIBYLL2.1 provides better, than QGSJET01, description of muon 
flux data up to several hundred GeVs~\cite{unger_ecrs2004} is not a basis to 
reduce all the problem to correct or incorrect choice of the EAS model. Our 
calculations show, that this model produces more positive, than negative, muons 
for small $E_\text{primary}/E_\text{threshold}$ ratio values both in showers 
from protons and neutrons, while for QGSJET01 and VENUS $N_{\mu^+}/N_{\mu^-}$ 
is less, than unity, in showers from primary neutrons. We also found, that 20\% 
difference between SIBYLL2.1 and QGSJET01 in total muon flux is almost entirely 
due to the difference in the flux of positive muons. This causes overestimation 
of muon charge ratio when one applies SIBYLL2.1~\cite{unger_ecrs2004}. As one 
can see, none of the current EAS models reproduces the data on muons, problems 
with description of the data on other EAS observables are briefly discussed 
in~\cite{watson2004}. By now, there remain large discrepancies between results 
on PCR energy spectra, extracted from the different EAS characteristics,  
indicating on disbalance in description of electromagnetic and hadronic 
components properties. It is necessary to add, that more definite conclusions 
on drawbacks of interaction models may be obtained if to apply them as well for 
processing of direct PCR spectra measurements~\cite{ya2004}.

\begin{figure}[bt]
\psfrag{Ryan et al.}[l][l]{Ryan et al.}
\psfrag{JACEE}[l][l]{JACEE}
\psfrag{MUBEE}[l][l]{MUBEE}
\psfrag{SOKOL}[l][l]{SOKOL}
\psfrag{RUNJOB2001}[l][l]{RUNJOB'01}
\psfrag{RUNJOB2004}[l][l]{RUNJOB'04}
\psfrag{KASCADE 2004}[l][l]{KASCADE (hadrons)}
\psfrag{KASCADE EM (QGSJET)}[l][l]{KASCADE $N_{e,\mu}$ (QGSJET)}
\psfrag{KASCADE EM (SIBYLL)}[l][l]{KASCADE $N_{e,\mu}$ (SIBYLL)}
\psfrag{Tibet (HD)}[l][l]{Tibet (HD)}
\psfrag{Tibet (PD)}[l][l]{Tibet (PD)}
\psfrag{1e2}{$10^2$}
\psfrag{1e3}{$10^3$}
\psfrag{1e4}{$10^4$}
\psfrag{1e5}{$10^5$}
\psfrag{1e6}{$10^6$}
\psfrag{1e7}{$10^7$}
\begin{center}
\includegraphics*[width=0.9\textwidth]{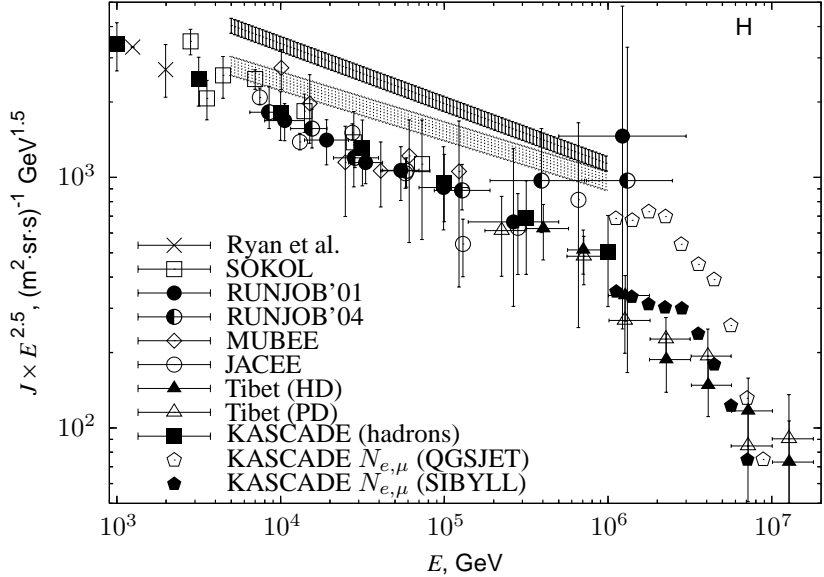}
\caption{Primary proton spectrum (complete list of references may be 
found in~\cite{ya2004,yaAGU2004_eng}). See explanation in text.} 
\end{center}
\end{figure}

Returning to the muon deficit problem one should not overlook existing 
uncertainties in the experimental data on muon intensity for $E_\mu>100$ GeV. 
They do not allow to give more precise estimates of discrepancy between 
calculated and measured fluxes. Fortunately, underground experiments provide 
the needed information for higher energies $E_\mu=1-10$ TeV. Reconstruction of 
muon spectrum at sea level from these data requires accurate description of 
muon transport in a dense medium. For this purpose we have applied a numerical 
method of adjoint equation solution and obtained muon intensities at large 
depths of rock and water with account of fluctuations in all muon interaction 
processes~\cite{yaAGU2004_zemlya_eng}. Our results are in good agreement with 
the results of Monte-Carlo codes MUM~\cite{mum} and MUSIC~\cite{music}. It is 
important to note, that our calculations give upper estimate of muon fluxes at 
large depths in comparison with MUM and MUSIC. This happens for the fact, that 
we used 1\% lower muon energy losses. From comparison of computed absorption 
curves with the data of underground installations we came to conclusion, that 
they are adequately described by the well-known muon spectrum~\cite{bugaev}. It 
provides good agreement with the data of LVD, KGF, Frejus collaborations and 
even underestimates data of MACRO, Soudan and BNO for depths up to 8 km~w.e., 
corresponding to $\sim10$~TeV median muon energy at sea level. Let us note, 
that muon intensity from~\cite{bugaev} exceeds intensity, obtained from direct 
data on PCR spectra with QGSJET01~\cite{ya2004} by $\sim45$\% in the energy 
range 1--10 TeV. In order to reproduce behavior of the spectrum~\cite{bugaev} 
for the given energies, we used interaction models SIBYLL2.1 and QGSJET01 with 
CORSIKA as EAS simulation code (for calculation procedure, see~\cite{ya2004}). 
As input information we applied PCR spectra parameterizations, proposed 
in~\cite{gaisser2002}. Since primary protons on $\sim70$\% determine muon flux 
at sea level, we have tuned their spectrum to match behavior of muon spectrum 
from~\cite{bugaev} within $\pm5$\% for $E_\mu=1-10$ TeV (corresponding primary 
energies are $5-10^3$ TeV). Formulae for heavier nuclei~\cite{gaisser2002} were 
taken without changes. The results, in comparison with available experimental 
data, are presented in Fig.~1. The upper shaded band is for QGSJET01: 
$J_p=(2.90-3.33)\times10^4E^{-2.74}$, and the lower one is for SIBYLL2.1: 
$J_p=(1.40-1.67)\times10^4E^{-2.70}$ (units are 
$(\text{m}^2\cdot\text{sr}\cdot\text{s}\cdot\text{GeV})^{-1}$). Spread in 
proton intensity for particular model reflects the uncertainty in the helium 
flux data according to~\cite{gaisser2002}. Before discussing reasons of the 
disagreement with the directly measured fluxes we should note, that obtained 
here proton spectra are not considered as the ``final'' versions: muon spectrum 
from~\cite{bugaev} may be reproduced by proton spectra with slightly different 
set of coefficients and power indexes (including energy depending ones) and 
possible underestimation of heavier nuclei fluxes cannot be excluded. The 
relevant to this situation result was recently presented by 
EAS-TOP/MACRO~\cite{eas_top_macro2004}. In this experiment primary 
\textit{p+He}\/ flux was derived with QGSJET01 from Cherenkov light integral spectra 
and radial distributions. Subtraction of proton component from the total 
\textit{p+He}\/ intensity gave twice larger, than obtained by JACEE, 
helium flux at the energy of 80 TeV (note, however, rather large systematic 
errors). The given result and the muon deficit problem provide enough evidences 
in favor of hypothesis, that light nuclei fluxes are systematically 
underestimated in the direct experiments. Discussion of methodical errors, 
which can be responsible for this, may be found 
elsewhere~\cite{runjob_eng,grigorov1973_eng,ya2004}. Additional information on 
this subject gives recent paper~\cite{kamae_gev_excess}, devoted to the 
galactic diffuse gamma-ray ``GeV excess'' problem. In this work it is shown, 
that account for Feynman scaling violation and diffractive interactions leads 
to 30--80\% increase of $\pi^0$'s, produced in $pp-$collisions, and the 
spectrum of incident protons is softer, than that of secondary $\gamma-$rays. 
Regarding the procedure, applied in the EC experiments, such effects would 
rather lead even to reduction of reconstructed PCR intensities (see, 
e.g.~\cite{ya2004,runjob_eng}, for more details). To make correct deduction on 
this question, first, it is necessary to evaluate the given effects for 
proton-nucleus, nucleus-nucleus collisions and their influence on cascade 
development in EC. Second, it should be accounted, that the scaling violation 
does not allow any more to get PCR spectrum from the electromagnetic cascades 
one with simple constant energy shift: at the least, the shift coefficient 
becomes energy dependent. And the third, it is necessary to estimate size of 
systematic errors, inevitably introduced in EC data by the use of 
semi-empirical models, relying on the validity of scaling hypothesis in 
extrapolation of low-energy and incomplete accelerator data to high-energy 
region. 

Though the modern EAS models incorporate scaling violation and diffractive 
interactions, none of them does it properly. This was demonstrated by KASCADE 
experiment on the basis of electromagnetic and hadronic events rate 
study~\cite{kascade_rates}. In particular, it was shown, that in QGSJET01 the 
fraction of diffractive dissociation in the total \textit{p-Air}\/ inelastic 
cross-section must be diminished by 6.5\% (i.e. halved). This is required to 
match the data on the observed hadronic events rate, which is 70\% lower, than 
calculated with QGSJET01~\cite{kascade_rates}. Such model modification would 
influence on the other KASCADE result~\cite{kascade_p2004}: primary proton 
spectrum, reconstructed from flux of single hadrons, reaching the ground (full 
squares in Fig.~1). Qualitatively it is clear, that \textit{larger}\/ primary 
$p$\/ flux would be needed to reproduce hadron spectrum, already not so 
perfectly conforming to the direct experiments data. The use for this purpose 
of SIBYLL2.1, where fraction of diffractive dissociation amounts to $\sim5\%$ 
at $10^4$~GeV and rapidly decreases to 2\% at $10^7$ GeV~\cite{luna_diffract}, 
can possibly lead even to larger increase of primary $p$ flux. Reduction of 
diffractive part of inelastic cross-section has another consequence for the 
muon deficit problem. It leads to the earlier shower development, hence, to 
higher probability of $\pi,K-$decays and to increase of muon number in EAS. For 
high-energy thresholds competitive process of muon decay can be neglected. Let 
us, however, note, that beside this factor, very important role in muon 
spectrum formation plays fraction of $\pi,K-$mesons, carrying the most part of 
primary particle energy. So, for high portion of diffractive events and high 
charged particle multiplicity, number of pions and kaons, falling into region 
$E_{\pi,K}/E_\text{primary}>0.1$, is smaller in QGSJET01, than in SIBYLL2.1. As 
a consequence, the latter model gives larger muon flux. Basing on the same 
arguments from available information on QGSJETII~\cite{engel_pylos} one may 
assume, that its use would bring to the intermediate, between SIBYLL2.1 and 
QGSJET01, values of muon flux. Finally, it can be concluded, that hardening of 
$\pi,K-$spectra and decrease of diffraction dissociation cross-section in 
QGSJET01 should result in better mutual agreement of primary proton spectra, 
reconstructed from hadron and muon fluxes. Notice also, that deduction on the 
need in harder, than in QGSJET01, spectra of secondary pions and kaons was also 
obtained in~\cite{kascade_rates} on the basis of hadron multiplicities 
examination. 

Another evidence of disbalance in description of hadronic and electromagnetic 
components also comes from KASCADE experiment~\cite{kascade_knee2004}. In the 
given paper, PCR energy spectra were reconstructed from electron-vs-muon number 
distribution. Proton spectra, taken by us from figures 
in~\cite{kascade_knee2004}, are shown in Fig.~1 with pentagons (error bars are 
omitted). It can be stated, that if to take into account QGSJET01 
modifications, proposed above, then all three spectra, derived from EAS 
observables (muons, hadrons, muons-vs-electrons) with this model, will be in 
satisfactory agreement. The use of SIBYLL2.1 leads to larger inconsistencies: 
it is evident, that application of $p$ flux from~\cite{kascade_knee2004} (full 
pentagons in Fig.~1) will enhance muon deficit. Let us note methodical aspect 
of this paper results: PCR energy spectra reconstruction procedure shows high 
sensitivity to the choice of hadronic generator, that is why it is required to 
perform such analysis in relation to the data, obtained in direct measurements. 

\section*{Concluding remarks}
Analysis of different kinds of EAS observations, performed in this paper, gave 
us evidences about possible underestimation of primary nucleon flux in direct 
experiments and information on drawbacks of QGSJET01 model (too soft 
$\pi,K-$spectra and high fraction of diffractive events). These conclusions 
hold rather qualitative character. We can not definitely say, that ``true'' 
primary proton spectrum lies between SIBYLL2.1 and QGSJET01 predictions, 
derived from the muon flux data. One cannot exclude, that significant part of 
primary nucleon flux underestimation is due to underestimation of nuclei fluxes 
with $Z\geq2$, which are subject to large systematic uncertainties. Correct 
energy dependence of diffraction cross-section and specific shape of secondary 
$\pi,K-$spectra in reggeon models also can hardly be given. It cannot be 
pointed out, which portion of the model modification relates to simple 
parameters tuning, and which to conceptual changes. To settle this questions, 
consistency of the interaction models must be checked together against data of 
direct and indirect (EAS) measurements, that suggests investigation of EC data 
sensitivity to variations of hadron--nucleus interaction characteristics. 

This work is supported in part by grants UR No.~02.01.001 and RFFI 
No.~04--02--16724.


\end{document}